# Replacing ANSI C with other modern programming languages

L. Dobrescu, *Member, IEEE*

*Abstract*—Replacing ANSI C language with other modern programming languages such as Python or Java may be an actual debate topic in technical universities. Researchers whose primary interests are not in programming area seem to prefer modern and higher level languages. Keeping standard language ANSI C as a primary tool for engineers and for microcontrollers programming, robotics and data acquisition courses is another strong different opinion trend. Function oriented versus object oriented languages may be another highlighted topic in actual debates.

*Index Terms*—Computer languages, Software engineering, Programming environments, ANSI C, Python, Java.

## I. INTRODUCTION

PROGRAMMING LANGUAGES are basic subjects in all technical and engineering universities. ANSI C, or simply C, first developed in 1969 at AT&T Bell Labs is worldwide known as a basic programming language. In fact it consists in many successive standards published by the American National Standards Institute (ANSI). Software developers writing in C always highlight its portability between compilers. C language is also the forefather of powerful supersets including C++ and Objective-C. The new developed programming language has primary the power of an object oriented language. C++ is a mid-level language, static typed, free-form, multi-paradigm and it supports procedural programming, but polymorphism is one of the prominent qualities of C++. It is as efficient and portable as C. It supports multiple programming styles (procedural programming, data abstraction, object-oriented programming, and generic programming).

ANSI C language, used in many engineering faculties as students´s aided design tool became a very modern and important subject regarding keeping it or not. It was used for many years in university environment, but there is a new wave to be replaced with other new and improved programming languages.

The newest Python programming language, introduced in 1991 is a high-level programming language, dynamic and efficient general-purpose. It is often used as a scripting language, but it is also a standalone program. Py2exe is software for writing programs in Python language and after that these can be packaged into executable programs. Its new multimedia features, its friendly and interactive interpreter and its flexibility provide a generally useful software tool, a testing and developing programming tool.

As microprocessors evolved and electronic devices increased in complexity and embedded system are currently available on the market, new hardware and new instructions has been designed for the new tasks. These microprocessors became known as CISC or Complex Instruction Set Computers.

C is a general-purpose programming language. It can work on any microprocessor with a designed C compiler. Moving easily between microcontroller families is a common skill of C programmers. They can write software faster and faster and they can create code that is much easier to understand and maintain [1].

Python and C are both powerful programming languages. They can be used for a variety of programming tasks. Each has benefits and drawbacks.

But the main difference between C and Python is that C uses a compiler, while Python uses an interpreter, as shown in Table I.

A major reason for keeping C over other languages in competitions in universities is its well known speed. When the program design and writing is timed, when the pressure of execution is important, C language ensures the right weapon.

The built in libraries or libraries developed by users are also good reasons for choosing C language.

Java can be another option for starting programming courses in universities. Learning the concepts of object-oriented programming is the most important option in choosing the right basic programming course.

The question about if one is better than the other seems to be normal, but finally not important and the answer depends on the application´s final goal. Many programmers are fluent in both C and Python or in both C and Java. But notice that all are fluent in C.

## II. DEBATES ON PYTHON

The main reason for replacing ANSI C programming language with Python stands in its complex syntax, and variables handling and management from a beginner point of view. Python seems to be more suitable as a starting point, as a first language in programming. Its airy and extremely easy

L. Dobrescu is with the University "Politehnica" of Bucharest and is the Vice Dean of Electronics, Telecommunications and Information Technology (e-mail: lidia.dobrescu@ electronica.pub.ro)

```
>>> x = int(raw_input("Please enter an integer: "))
Please enter an integer: 42
>>> if x < 0:
...     x = 0
...     print 'Negative changed to zero'
... elif x == 0:
...     print 'Zero'
... elif x == 1:
...     print 'Single'
... else:
...     print 'More'
...
More
```

Fig. 1. Python decision structure.

syntax encourage a beginner programmer in a continuously study, tests and experimenting.

Python has efficient high-level data structures, as shown in figure 1 and a simple approach towards object-oriented programming.

Its syntax is elegant and this enables a dynamic typing, its interpreted nature develops an ideal language for scripting and rapid application [2].

Every C course starts with variables definitions, types, modifiers, classifications and a lot of tables explaining variable ranges and limits.

Python syntax is quite similar to C language. It allows the programmers to use a code also like C language. Scripting languages provide an interactive and dynamic environment using web-based applications, as shown in figure 2.

Python can offer an amazing experience at a first glance, inviting to persevere in learning new things and this experience can be very important even for older people with no programming experience at young age.

The ability to program on the Web has became critical to many professions and students should have Web frameworks available after a programming course.

In Python, the ability to program desktop applications while trends are moving towards the web can be seen as another strong point.

Another facilities such as a marketable professional skills and a supportive and welcoming community around the new Python programming language are also strong points in replacing ANSI C with Python.

```
C Fragment                      Python Equivalent

disc = b * b - 4 * a * c;       disc = b * b - 4 * a * c
if (disc < 0)                   if disc < 0:
{                                   num_sol = 0
    num_sol = 0;                else:
}                                   t0 = -b / a
else                                if disc == 0:
{                                       num_sol = 1
    t0 = -b / a;                        sol0 = t0 / 2
    if (disc == 0)                  else:
    {                                   num_sol = 2
        num_sol = 1;                    t1 = disc ** 0.5 / a
        sol0 = t0 / 2;                  sol0 = (t0 + t1) / 2
    }                                   sol1 = (t0 - t1) / 2
    else
    {
        num_sol = 2;
        t1 = sqrt(disc) / a;
        sol0 = (t0 + t1) / 2;
        sol1 = (t0 - t1) / 2;
    }
}
```

Fig. 2. C versus Python demonstrative program

TABLE I
COMPILER VERSUS INTERPRETER

| No | Compiler | Interpreter |
|---|---|---|
| 1 | The entire program is taken as an input. | Takes Single instruction as input. |
| 2 | Intermediate Object Code is generated. | No Intermediate Object Code is generated. |
| 3 | Conditional Control statements are executed faster. | Conditional Control statements are executed slower. |
| 4 | More memory is required since Object Code is generated. | Memory requirements is less. |
| 5 | Program need not be compiled every time. | Every time higher level program is converted into lower level program. |
| 6 | Errors are displayed after entire program is checked. | Errors are displayed for every instruction interpreted. |
| EXAMPLE | Python, Ruby | C, C++ |

Simple facts in programming such as using Python as a calculator, Python keywords like and, or, not instead of &&,|| and !, the bracelets 'absence can simplify many problems. The Python loops and vectors are also simplified.

The error checking in Python is also better than in ANSI C. As a high-level language, Python has high-level data types built in, such as flexible arrays and dictionaries as another major advantage.

Even Python uses an interpreter, the source code is compiled into bytecode and transformed in another files with pyc or pyo extensions. Executing such files becomes faster by avoiding recompilation. This manner is also called a virtual machine that executes the bytecode. The bytecodes usually do not run on different Python virtual machines and are not stable running on many different Python releases.

After learning and programming in Python, the migration towards C environment seems to be much more natural.

Many young students enrolled in basic programming courses start by learning a procedural and statically typed language. Examples can include C, Pascal, or Java.

There are many strong points to learn these languages, but from a beginner point of view Python enables developing programming skills free of syntax constraints. More like Java, Python's large standards library allows easily creating functional programs.

III. DEBATES ON JAVA

The things become more complicated considering Java programming language.

Java is a newer developed software and mainly it allows a procedural programming. Since Java 5, it enables a generic programming. Since Java 8, the new software enables a functional programming. The main characteristic of Java is the strongly he object-oriented programming paradigm. It includes support for the creation of scripting languages.
There are many debating plans regarding Java.

The first of them starts from object oriented programming technique as a fundamental environment.

If a new student has never used an object-oriented programming language before, then basic concepts must be learned before he can begin writing any code.

The Java first lessons will introduce him to objects, classes, inheritance, interfaces, and packages.

Many discussions focuses on how these concepts are related to the real world, while simultaneously providing an introduction to the complex syntax of Java programming language.

A very modern and original second debating plan stands between Python and Java.

It is generally known that Python runs slower than Java programs, but it takes much less time to develop a new program. Python programs are generally many times shorter than the same Java programs. Python's built-in high-level data types and its dynamic typing can explain these .

In a Python program the types of arguments and variables are simply not declared. Python's has a powerful polymorphic list and dictionary types and its rich syntactic support is built in from the beginning.

A classical example evaluates the sum of two variables, generally named a and b. Java must first inspect the objects a and b and to establish type. After that the suitable addition operation is performed, which may be an overloaded user-defined method. An efficient integer or floating point addition can be performed, but this requires variable declarations for a and b. It does not allow overloading of the sum operator for instances of user-defined classes.

Python is much better suited as a "glue" language, while Java is better characterized as a low-level implementation language.

In fact, an excellent mixture can be done from them.

```
int main (void)
{
    DDRD |= 1<<PD6;   //output pins
    DDRD |= 1<<PD4;

    uint8_t press_key;

    ioinit();

    while(1)
    {
        press_key = uart_receive-character ();
    //red LED
    if(press_key == 'b')
        {printf("ON\n");
        PORTD |= (1<< PD4);}
    if(press_key == 'z')
        {printf("OFF\n");
        PORTD &= ~(1<<PD4);}
    if((press_key == 'Y') && (PORTD &= (1<<PD4)))
        printf("on\n");
    else if((press_key == 'Y') && (!(PORTD &= (1<<PD4))))
        printf("off \n");

    //green LED
    if(press_key == 'a')
        {printf("ON\n");
        PORTD |= (1<< PD6);}
    if(press_key == 's')
        {printf("OFF\n");
        PORTD &= ~(1<<PD6);}
    if((press_key == 'X') && (PORTD &= (1<<PD6)))
        printf("ON\n");
    else if((press_key == 'X') && (!(PORTD &= (1<<PD6))))
        printf("OFF \n");
    }
    return(0);
}
```

Fig. 3. ATmega16 program

Components can be developed in Java and combined to form applications in Python; Python can also be used to prototype components until their design can be "hardened" in a Java implementation.

The third debating plan discourages Java introduction as a first programming language.

Compared to a compiled C++ program, Java has low performance. The JV Machine is often unreliable because its frequently updating .

Java is a definitely Object Oriented programming paradigm. Everything must be clear defined in terms of objects and classes, and even very simple operations such as the well known "Hello World!" program must be written on several lines of code.

In C++, the student can switch between imperative and Object Oriented paradigm, more easily.

## IV. DEBATES ON KEEPING ANSI C

On the other hand one must consider the fact that, in electronics faculties, the microcontrollers programming and data acquisition systems are standard courses for engineering students and lately the projects practical implementations on boards with microcontrollers such as ARDUINO are using ANSI C programming language.

AVR is RISC architecture and its instruction set based on C programming language. C compiler helped the hardware design in order to optimize it for C programming.

The ATmega16 is a 8-bit AVR RISC, low-power CMOS microcontroller. The AVR microcontroller was connected with PC. In the main program, shown in figure 3, PD6 and PD4 have been defined as the pins where two LEDs, a green and a red one were connected as output loads. A new variable press_key was defined to keep the input character. An init function for the serial link was defined. An infinite while was created for input character. The „b" character for lighting and „z" for stop and „Y"is for interrogation. ANSIC is a widely used programming language.

The majority of computer architectures and operating systems use ANSI C. This programming language has also tools for structured programming.

The distinguish feature of a structured language is compartmentalization of code and data. This property allows C programmers to share sections of code.

The end result is that C implements few restrictions, few complaints, block structures, stand alone functions and a compact set of keywords [3].

Many later languages are C related and inherit a lot of its syntax elements such as C#, Objective-C, Perl, PHP, Python, and even Java or Java Script.

They have drawn many of their control structures from ANSI C language.

C was designed for low-level memory management tasks. Pointers are the main powerful tool for memory access. C is at the same time a low level language because it has a direct access to memory locations but it is also a high level language because it allows a structured mode of user interface with the computer. Its unique facility of using pointers is the most significant weapon of this language. Only good programmers feel the power of this important C tool.

Usually during high school, students start learning C language, usually Borland C version. The better debugger of Borland C is an important reason for starting with it.

After learning the first steps in the new programming environment, in the first year in engineering universities, the students begin to use ANSI C.

It may seem a regressive way, but in fact they learn to use the basic instruction set of C that ensures the real portability of written source codes. Using printf and scanf input-output instructions, the future engineers begin to learn the first instructions for microcontrollers programming.

C has been standardized and improved throughout the years. It can be seen as the ancestor of C++ and Objective-C and the newest applications for Apple's IOS written exclusively in Objective-C.

Teaching methods of C programming became lately an important subject of debate for teachers, because these must be improved in order to capture the student's attention and not to transform a programming course or any course in a boring lecture. It's urging to abandon chalk, using modern methods, eventually it can be used to write only to highlight mistakes, short examples or new hints on the blackboard.

Lately more and more teachers prefer the "Power Point method".

Although it seems to be a modern method, very comfortable for teachers, it isn't so helpful in teaching programming languages because it ends in overwhelming and boring students. PowerPoint is now widely used in lectures to science students in most colleges of China.

Some of its advantages are producing better visual effects, high efficiency in information transfer, precise and systemic knowledge structure. Disadvantages of PowerPoint may be induced by irrelevant information in slides, neglect of interaction with students, uncontrolled speed in presenting or too strict order of slides [4].

Is not necessary to remove or replace this method of teaching C, but it must be improved and updated. Ideally, each student must have a laptop connected to a smart board and when teacher gives to all students to solve a programming problem, this would be displayed on board with the solution. The solution must be chosen among the first well solved problems on student's laptops. Using graphs, colored pictures and diagrams, figures, the teacher's lecturers may become more attractive.

In figure 4, a colored example of C program structure is shown as it is used in a ANSI C Course for students [5].

C++ is compiled directly into an executable file that can run in almost any situation as long as it remains on the same operating system.

In the picture, important parts of a program are described using different colors.

An instructor may use a different mix of technologies in the classroom and use them creatively in order to promote the learning of students and to satisfy students' learning needs and objectives [6].

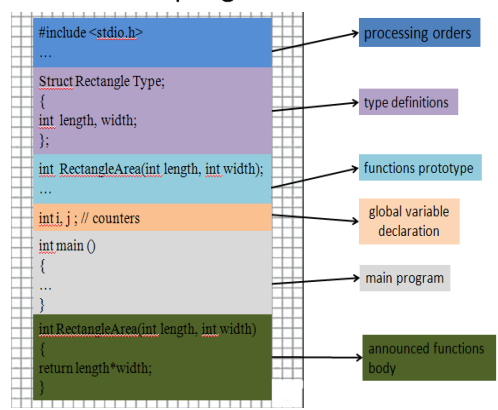

Fig. 3. ATmega16 program

By onscreen synchrony of PowerPoint slides and recorded voice, so-called "e-lecture" may be delivered on web and serve as a useful alternation of traditional lecture, either alone or combined with other methods [7].

V. CONCLUSIONS

Replacing ANSI C language with other modern programming languages represents a modern wide debate topic.

ANSI C is a general-purpose programming language. Programming microcontrollers, robotics and data acquisition courses using ANSI C, imposes this language in any technical university.

Python can be more suitable as a first language in programming. It is compact and comprehensible. As a modern high-level language it also has high-level data types built in such as flexible arrays and dictionaries.

The main characteristic of Java is the strongly object-oriented programming paradigm. Its basic concepts including objects, classes, inheritance, interfaces, and packages must be learned before writing any code.

Many young students enrolled in basic programming courses start learning ANSI C and then Java.

Many programmers are fluent in both C and Python or in both C and Java.